\documentclass{nature}
\usepackage{graphicx}
\usepackage{bm} 
\usepackage{pifont}
\usepackage{amsmath, amsthm, amssymb, pifont, wasysym}
\usepackage[labelfont=bf]{caption}
\DeclareCaptionLabelSeparator{bar}{$|$\ }
\usepackage[labelsep=bar]{caption}
\usepackage{indentfirst}
\usepackage{color}
\usepackage{dcolumn} 
\usepackage{siunitx} 
\DeclareSIUnit{\ohm}{\ensuremath{\Omega}} 
\usepackage{braket}

\newcommand {\SRO}{SrRuO$_3$}
\newcommand {\STO}{SrTiO$_3$}

\newcommand {\TC}{$T_{\mathrm{C}}$}
\newcommand {\rhoxx}{${\rho_{\mathrm{xx}}}$}
\newcommand {\rhoyx}{${\rho_{\mathrm{yx}}}$}

\newcommand {\Syx}{$S_\mathrm{yx}$}

\newcommand {\SyxConeB}{$S_{\mathrm{yx,}C_\mathrm{1,}B}$}
\newcommand {\SyxConeT}{$S_{\mathrm{yx,}C_\mathrm{1,}T}$}

\newcommand {\SyxzeroT}{$S_{\mathrm{yx},\text{\SI{0}{T}}}$}
\newcommand {\Sxx}{$S_\mathrm{xx}$}

\newcommand {\Vy}{$V_\mathrm{y}$}
\newcommand {\dxT}{$\varDelta _\mathrm{x}T$}
\newcommand {\nablaxT}{$\nabla _\mathrm{x}T$}
\newcommand {\axy}{$\alpha_\mathrm{xy}$}
\newcommand {\Morb}{$M_\mathrm{z}^\mathrm{orb}$}
\newcommand {\dMorb}{$\Delta M_\mathrm{z}^\mathrm{orb} (\mu)$}

\title{{\color{black}Observation of in-plane anomalous Nernst effect}}

\author{Tadashi Yoneda$^{1}$, Shinichi Nishihaya$^{1}$, Markus Kriener$^{2}$, Haruto Kaminakamura$^{1}$,\\
{\color{black}Ming-Chun Jiang$^{2}$, Naohiro Tezuka$^{1}$,} Yoshiya Murakami$^{1}$, {\color{black}Ryotaro Arita$^{2,3}$,} Hiroaki Ishizuka$^{1}$,\\
Masaki Uchida$^{1,4\ast}$}

\begin{document}

\maketitle
\renewcommand{\baselinestretch}{1.5}\normalsize
\begin{affiliations}
\item Department of Physics, Institute of Science Tokyo, Tokyo 152-8551, Japan
\item RIKEN Center for Emergent Matter Science (CEMS), Wako 351-0198, Japan
{\color{black}\item Department of Physics, The University of Tokyo, Tokyo 113-0033, Japan}
\item Toyota Physical and Chemical Research Institute, Nagakute 480-1192, Japan
\end{affiliations}

\renewcommand{\baselinestretch}{2}\normalsize
\begin{abstract}
The Nernst effect, which enables the conversion of a heat current into a transverse voltage under magnetic field or spin magnetization, holds significant promise for energy harvesting and thermal management in future electronics. However, the conventional Nernst effect is fundamentally constrained by the orthogonality requirement that the applied field or spontaneous magnetization must be perpendicular to the plane defined by the temperature gradient and the induced voltage. Here we report that symmetry-tailored ultrathin films of a prototypical ferromagnetic oxide exhibit anomalous Nernst effect arising from intrinsic coupling to spontaneous in-plane spin magnetization. Systematic magnetothermoelectric measurements under spherical rotations of the magnetic field reveal that a pronounced Nernst signal, comparable in magnitude to the out-of-plane response, emerges robustly associated with out-of-plane orbital magnetization. Our findings demonstrate that the anomalous Nernst effect is no longer limited by the orthogonality condition, opening new opportunities for more flexible designs of magnetothermoelectric materials and devices.
\end{abstract}
\newpage

Thermoelectric phenomena, which enable the direct conversion between heat and electricity, have long been explored in the solid-state physics, driven by prospects for energy harvesting and thermal management. 
The Nernst effect, one of the magnetothermoelectric phenomena, refers to the generation of a transverse voltage perpendicular to both the temperature gradient and the applied magnetic field, arising from the Lorentz force acting on charge carriers \cite{firstNE}. 
In materials with magnetization, such as ferromagnets, an additional contribution emerges, known as the anomalous Nernst effect (ANE) \cite{firstANE}. 
{\color{black}In the experimental search for materials exhibiting large Nernst response for device applications, most previous investigations of ANE have primarily focused on identifying compounds with large spontaneous spin magnetization perpendicular to the plane defined by the temperature gradient and the induced voltage, as illustrated in Fig. 1\textbf{a} \cite{AHEandANE1, AHEandANE2, AHEandANE3, ANEexp1, ANEexp2, ANEexp3, ANEexp4, AHEandANEcal1, ANEexp5, ANEexp6, ANEexp7, ANEthermopile}. 
More recently, based on theoretical formulations of ANE in terms of the Berry curvature, experimental efforts have increasingly focused on characteristic band structures with large Berry curvature, rather than simply on materials with large spin magnetization \cite{ANEexp1, ANEexp2, ANEexp3, ANEexp4, AHEandANEcal1, ANEexp5, ANEexp6, ANEexp7, ANEthermopile}.}
This has yielded significant conversion efficiencies, particularly in magnetic materials hosting Berry curvature hot spots such as band crossing points and lines \cite{ANEexp1, ANEexp2, ANEexp3, ANEexp4, AHEandANEcal1, ANEexp5, ANEexp6, ANEexp7, ANEthermopile}. 
However, the requirement that the spin magnetization, temperature gradient, and induced voltage must be mutually orthogonal remains unchanged, potentially restricting the flexibility of device layouts in practical applications.

In the case of anomalous Hall effect (AHE), which typically manifests as a transverse voltage in response to an electric current in the presence of out-of-plane spin magnetization \cite{AHEreview}, this restriction is gradually being overcome \cite{iAHEtheexp_ECS, iAHEtheexp_Fe3Sn2, iAHEthe1_Q, iAHEthe3_Q, iAHEthe_otherconv1, iAHEthe4_Q, iAHEthe6, iAHEthe2_Q, iAHEthe_otherexo1, iAHEthe_otherexo2, iAHEthe8, iAHEthe7_sym, iAHEsym, multipolar, iAHEexp_Fe, iAHEexp_Co3Sn2S2, iAHEexp_EZS, iAHEexp_Cd3As2, SRO_iAHE}. Specifically, the emergence of a transverse voltage under spin magnetization in the current-voltage plane has been demonstrated on the principal plane, exhibiting three-fold rotational symmetry with respect to in-plane field rotation \cite{iAHEtheexp_ECS, iAHEtheexp_Fe3Sn2,iAHEexp_Fe,iAHEexp_Co3Sn2S2,iAHEexp_EZS,iAHEexp_Cd3As2,SRO_iAHE}, clearly distinct from the conventional ordinary and anomalous Hall effects. One possible origin lies in the modulation of band crossing points by the in-plane magnetic field \cite{iAHEtheexp_ECS}, through a spin-orbit coupling term between the in-plane momentum and the out-of-plane spin operator \cite{iAHEtheexp_ECS, iAHEthe1_Q, iAHEthe3_Q, iAHEthe_otherconv1, iAHEthe4_Q, iAHEthe6}, which gives rise to off-diagonal coupling between in-plane spin magnetic field and out-of-plane orbital magnetization \cite{orbitalmag_rev, orbitalmag_thermo, orbitalmag1, orbitalmag2, orbitalmagterm1, orbitalmagterm2}. However, it is highly nontrivial whether the same mechanism can be responsible for ANE, because ANE is primarily governed by the Berry curvature at the Fermi energy \cite{orbitalmag_rev, orbitalmag_thermo}, whereas the intrinsic AHE involves its integration over all occupied bands \cite{AHEreview}. How the energy derivative appearing in Mott's formula relates ANE and AHE is an intriguing problem particularly in more exotic mechanisms \cite{iAHEtheexp_Fe3Sn2, iAHEthe2_Q, iAHEthe_otherexo1, iAHEthe_otherexo2, iAHEthe8}, even though the symmetry required for both is expected to be the same \cite{iAHEthe7_sym, iAHEsym, multipolar}.

Here we report the experimental observation of {\color{black}in-plane ANE, which arises under unconventional in-plane magnetic field lying parallel to the plane defined by the temperature gradient and the induced voltage, as illustrated in Fig. 1\textbf{b}.}
By systematically measuring ultrathin films of a typical ferromagnetic metal {\SRO} for various field directions, we demonstrate that a substantial {\color{black}in-plane} ANE signal emerges with coupling to the spontaneous in-plane spin magnetization.
Our findings broaden the potential of ANE by lifting the conventional orthogonality restriction among spin magnetization, temperature gradient, and generated voltage.

\section*{{\color{black}In-plane Anomalous Nernst effect}}
Magnetothermoelectric as well as magnetotransport measurements under magnetic field rotation were done by constructing the experimental setup shown in Figs. 2\textbf{a} and 2\textbf{b} (see Methods and Supplementary Note 1 for details). 
(111) {\SRO} films grown on the (111) {\STO} substrate are trigonally distorted with compressive epitaxial strain \cite{SRO_iAHE, SRO111_2}, hosting multiple Weyl points in the electronic band structure \cite{AHEexp1_SROWeyl1, SROWeylcal,SRO_iAHE,SRO111_1} and also satisfying symmetry requirements for the in-plane ANE. 
{\color{black}This symmetry requirements mean that {\SRO} with spin magnetization along [11$\bar{\text{2}}$] direction belongs to the magnetic point group $2'/m'$, in which the Berry curvature $\Omega_\mathrm{z}$ does not cancel when integrated over the Brillouin zone and allows a finite response tensor \cite{iAHEthe7_sym,iAHEsym,multipolar,tensorsym}.}
Figures 2\textbf{c} and 2\textbf{d} present the conventional Nernst signal {\Syx} and the Hall resistivity {\rhoyx}, measured with sweeping the out-of-plane magnetic field for a (111) {\SRO} thin film at the nominal sample temperature of 63 K. 
The Curie temperature {\TC} of this sample is determined to be \SI{130}{K} (see  Supplementary {\color{black} Figs. 3 and 4} for additional fundamental transport {\color{black}and magnetization data}) \cite{SRO111_1,SRO111_2, SRO_iAHE}. 
{\Syx} exhibits a small but finite hysteresis loop, indicative of a hard ferromagnetic behavior, and increases almost linearly above the coercive field with a slope or Nernst coefficient of \SI{0.012}{\micro V /KT}. 
{\rhoyx} exhibits a similar hysteresis loop, with a pronounced nonlinearity beyond the coercive field. 
Importantly, the signs of {\Syx} and {\rhoyx} loops, reflecting their respective anomalous components, are opposite. 
This sign relation is consistent with previous studies on (001) {\SRO} films \cite{SRO_ANE}, where it is argued that ANE is significantly influenced by modulation of Weyl points and Berry curvature by epitaxial strain. 
Since the present (111) {\SRO} films are compressively strained, the observed sign relation between ANE and AHE may be interpreted within the same framework.

Figures 2\textbf{e} and 2\textbf{f} compare {\Syx} and {\rhoyx} taken with sweeping the in-plane magnetic field at $\varphi = 0^\circ$ and $60^\circ$, where the azimuthal angle $\varphi$ is measured from the [11$\bar{\text{2}}$] direction. 
Here, in-plane field dependence of {\Syx} is obtained by averaging {\Syx} measured for three equivalent directions ($\varphi = 0^\circ, 120^\circ, 240^\circ$ or $\varphi = 60^\circ, 180^\circ, 300^\circ$) to reduce the one-fold symmetric component with respect to the in-plane field rotation (see Supplementary Note {\color{black}6} for details). 
Although such a one-fold symmetric signal has been previously reported in a different system \cite{iANEexp}, it is experimentally challenging to distinguish it from the conventional out-of-plane ANE signal, originating from possible field misalignment{\color{black}, out-of-plane temperature gradient,} and other extrinsic contributions in the magnetothermoelectric measurements. 
{\Syx} induced by the in-plane field exhibits significantly large values comparable to those observed under the out-of-plane field. 
It also remains finite at zero magnetic field accompanied by a large hysteresis, and varies slightly with increasing field above the coercive field. 
{\rhoyx} behaves similarly to {\Syx}, but with an opposite sign and a stronger field dependence, as in the out-of-plane field scan. 
These differences in the field dependence can be interpreted in terms of Mott's formula that relates the Nernst signal and the energy derivative of the Hall conductivity.

{\Syx} taken for $\varphi = 60^\circ$ exhibits a similar in-plane field dependence, but with a sign opposite to that observed for $\varphi = 0^\circ$. 
These observations of the in-plane ANE are consistent with the symmetry of the trigonally distorted (111) {\SRO} thin films, which feature a $C_3$ axis along [111], $C_2$ axes along {\color{black}[$\bar{\text{1}}$10]} and its equivalents, and mirror planes on {\color{black}($\bar{\text{1}}$10)} and equivalent planes as shown in Fig. 2\textbf{b}. 
Such symmetry allows the emergence of in-plane ANE with opposite signs, centered at $\varphi = 0^\circ , 120^\circ , 240^\circ $ and $\varphi = 60^\circ , 180^\circ , 300^\circ $ respectively. 
Importantly, the grown {\SRO} ultrathin films possess in-plane easy axes of spin magnetization {\color{black} realized on the delicate balance of competing in-plane shape magnetic anisotropy and $\langle \text{111} \rangle$ magnetocrystalline anisotropy\cite{SRO_iAHE}, as also confirmed by thickness-dependent in-plane AHE (see Supplementary Note 5 for details).}
{\color{black}Namely, in the ultrathin film, the out-of-plane spin magnetization component favored by the magnetocrystalline anisotropy is absent and a finite ANE emerges in the in-plane spin magnetization state aligned with the in-plane magnetic field.}
As another feature, a hump structure appears both in {\Syx} and {\rhoyx} after crossing zero field. 
This could be explained by superposition of inhomogeneous magnetic domains with different coercive fields, as examined for the out-of-plane AHE in {\SRO} ultrathin films \cite{twoAHE1}. 
However, the present hump structure is observed only for the in-plane field sweep, suggesting that the spin magnetization may be oriented away from the in-plane direction {\color{black}only} during its reversal.
{\color{black} One possible interpretation is that}, in the case of [11$\bar{\text{2}}$] spin magnetization, upon reversing the in-plane magnetic field, {\color{black}the spin magnetization may switch} from the [11$\bar{\text{2}}$] direction to the [$\bar{\text{1}}\bar{\text{1}}$2] direction via the [$\bar{\text{1}}\bar{\text{1}}\bar{\text{1}}$] direction due to the $\langle \text{111} \rangle$ magnetocrystalline anisotropy, rather than switching directly.
{\color{black} The detailed in-plane magnetization reversal process will be the subject of future studies.
}

\section*{Investigation by spherical field rotations}
Figure 3\textbf{a} shows the azimuthal angle dependence of {\SyxzeroT}, the Nernst signal remaining at zero magnetic field after the in-plane field sweep. 
Here {\SyxzeroT} is obtained by antisymmetrization of a pair of raw data $S_{\mathrm{yx},\text{\SI{0}{T}},\text{raw}}(\varphi)$ and $S_{\mathrm{yx},\text{\SI{0}{T}},\text{raw}}(\varphi + 180^\circ)$, as expressed by $S_{\mathrm{yx,}\text{\SI{0}{T}}}(\varphi) = (S_{\mathrm{yx,}\text{\SI{0}{T}}\mathrm{,raw}}(\varphi) - S_{\mathrm{yx,}\text{\SI{0}{T}}\mathrm{,raw}}(\varphi + 180^\circ))/2$. 
$S_{\mathrm{yx,}\text{\SI{0}{T}}}$ exhibits a three-fold symmetric change upon in-plane rotation of the polarizing field, consistent with the data taken by in-plane field sweeps at $\varphi = 0^\circ$ and $60^\circ$.

As confirmed in Figs. 3\textbf{b} and 3\textbf{c}, above the coercive field, {\Syx} as well as {\rhoyx} varies with three-fold symmetry for the in-plane field rotation. 
Here, {\Syx $(\varphi)$} and {\rhoyx $(\varphi)$} are obtained by the same antisymmetrization as {\SyxzeroT $(\varphi)$} and subsequent subtraction of the {\color{black}two types of one-fold components {\SyxConeB} and {\SyxConeT}.
{\SyxConeB} is caused by the out-of-plane field component due to misalignment between the field-rotation and transport measurement planes with a maximum at a particular azimuthal angle of the tilted direction. 
On the other hand, {\SyxConeT} arises from the out-of-plane temperature gradient component --- this needs to be considered in the in-plane Nernst measurement, showing a maximum at $\varphi = 0^\circ$ (see Supplementary Fig. 6).
By subtracting these extrinsic one-fold components, the in-plane ANE can be evaluated more accurately.}
{\Syx} exhibits a square-wave-like behavior at \SI{5}{T}, in contrast to the sinusoidal behavior at \SI{9}{T}.
This square-wave-like behavior reflects the in-plane anisotropy preferring [11$\bar{\text{2}}$] direction and equivalents, as more prominently observed in {\SyxzeroT}. 
A similar trend, though weaker, can also be seen in {\rhoyx}.

We have also examined the polar angle dependence of {\SyxzeroT} within {\color{black}($\bar{\text{1}}$10)} plane, which is obtained by antisymmetrization of a pair of $S_{\mathrm{yx},\text{\SI{0}{T}},\text{raw}}(\theta)$ and $S_{\mathrm{yx},\text{\SI{0}{T}},\text{raw}}(\theta + 180^\circ)$. 
As confirmed in Fig. 3\textbf{d}, {\SyxzeroT} remains finite even for the [11$\bar{\text{2}}$] in-plane direction ($\theta = 90^\circ$), with a value roughly $\SI{40}{\%}$ of the out-of-plane value ($\theta = 0^\circ$ and $180^\circ$) {\color{black}(see also Supplementary Fig. 9)}. 
Importantly, {\SyxzeroT} exhibits a plateau structure centered at the in-plane direction as well as the out-of-plane ones, reflecting the magnetic anisotropy. 
This magnetic anisotropy, favoring the [11$\bar{\text{2}}$] in-plane direction, is determined by the combination of the in-plane shape magnetic anisotropy {\color{black}enhanced in ultrathin films} and the $\langle \text{111} \rangle$ magnetocrystalline anisotropy. 
Figures 3\textbf{e} and 3\textbf{f} show $\theta$ dependence of {\Syx} and {\rhoyx}, taken upon rotating the field direction at \SI{5}{T} and \SI{9}{T}. 
$S_\mathrm{yx} (\theta)$ and $\rho_\mathrm{yx} (\theta)$ are obtained by the same antisymmetrization used for {\SyxzeroT $(\theta)$}. 
At \SI{5}{T}, {\Syx} as well as {\rhoyx} exhibits rather plateau like structures similar to one observed for {\SyxzeroT}.
{\color{black} Here, the contribution of {\SyxConeB} does not appear for $\theta = 90^\circ$ since the origin of $\theta$ is chosen to give the maximum out-of-plane component, and the contribution of {\SyxConeT} is subtracted by using the amplitude of {\SyxConeT} obtained in the in-plane field rotation measurement (see Supplementary Note 6 for details).
As confirmed in Fig. 3\textbf{e}, the value of {\Syx} at $\theta = 90^\circ$, under fully in-plane field configuration, is consistent with the value obtained at $\varphi = 0^\circ$ for the in-plane field rotation shown in Fig. 3\textbf{b}.
This observation of in-plane ANE, where the magnetic field and spin magnetization lie within the temperature gradient-voltage plane, demonstrate that ANE is no longer constrained by the conventional orthogonality condition.
}

\color{black}
\section*{Temperature dependence of in-plane ANE}

Figure 4\textbf{a} shows the in-plane field dependence of the transverse thermoelectric conductivity {\axy} at typical temperatures.
Here, {\axy} is obtained by the equation {$\alpha_{\mathrm{xy}}= (S_{\mathrm{yx}} + \theta_{\mathrm{AHE}}\cdot S_\mathrm{xx})/\rho_{\mathrm{xx}}$}, where {\Sxx} is the Seebeck coefficient, {\rhoxx} is the longitudinal resistivity, and $\theta_{\mathrm{AHE}} = \rho_{\mathrm{yx}}/\rho_{\mathrm{xx}}$ is the anomalous Hall angle.
Below {\TC}, {\axy} becomes finite and exhibits hysteresis with an increasing coercive field as the temperature decreases.
Also, the first term originating from {\Syx} is confirmed to be dominant in {\axy} (see Supplementary Fig. 10).
This indicates that the observed in-plane ANE is closely related to the ferromagnetic ordering in {\SRO}, specifically arising from the off-diagonal coupling between the in-plane spin magnetization and out-of-plane orbital magnetization as discussed later.

A significant difference between in-plane ANE and in-plane AHE is confirmed in their temperature dependences.
As compared in Fig. 4\textbf{b}, while {\rhoyx} monotonically increases as the temperature decrease, {$|\alpha_{\mathrm{xy}}|$} shows a maximum around \SI{80}{K} and then decreases toward zero.
This reflects the fact that the Nernst effect is governed by energy derivative of the Hall conductivity, leading to a linear suppression with temperature in the low-temperature limit as expressed by the Mott relation.
These results suggest that the present observation of in-plane ANE offers an alternative route to investigating Berry-curvature-related quantum geometry beyond in-plane AHE.
\color{black}

\section*{Discussion}
These magnetothermoelectric measurements performed for various magnetic field directions on the sphere confirm that the ANE and its spontaneous response indeed appear for the in-plane spin magnetization state. 
Even when the spin magnetization {\color{black}$\bm{M}^\mathrm{spin}$} lies entirely in-plane, the emergence of ANE indicates that the net magnetic dipole moment necessarily possesses an out-of-plane component. 
Namely, orbital magnetization, generated by the rotational motion of electron wave packets, manifests as the out-of-plane component of the net magnetization, playing a crucial role in the emergence of the in-plane ANE. 
\color{black}
Here, the orbital magnetization $\bm{M}^\mathrm{orb}$ and thermoelectric conductivity {\axy} are intrinsic properties related to the band structure, similar to the Berry curvature $\bm{\Omega}$ and anomalous Hall conductivity $\sigma_\mathrm{xy}$.
The orbital magnetization $\bm{M}^\mathrm{orb}$ is formulated as
\begin{equation}
  \bm{M}^\mathrm{orb} = \frac{e}{\hbar}\frac{1}{N_k V_c}\sum_{n\bm{k}}\frac{1}{2}f_{n\bm{k}}\operatorname{Im} \Braket{\nabla_{\bm{k}} u_{n\bm{k}}|H_{\bm{k}} + \varepsilon_{n\bm{k}} - 2\varepsilon_\mathrm{F}|\nabla_{\bm{k}} u_{n\bm{k}}},
\end{equation}
where $N_k$ is the number of $k$-points sampled in the Brillouin zone, $V_c$ is the cell volume, $f_{n\bm{k}}$ is the occupation number, $u_{n\bm{k}}$ is the Bloch wave function of band index $n$ and crystal momentum $k$ and $\varepsilon_\mathrm{F}$ is the Fermi energy \cite{orbitalmag_rev,orbitalmag1,orbitalmag2}.
The orbital magnetization {$\bm{M}^\mathrm{orb}$} can be decomposed into two contributions: one gives the local orbital moment, while the other is related to the Berry curvature and represents the itinerant, anomalous contribution {$\Delta \bm{M}^\mathrm{orb} (\mu)$}, with the chemical potential $\mu$.
The itinerant part of the out-of-plane orbital magnetization {\Morb} is directly connected to {$\sigma_\mathrm{xy}$} through the St\v{r}eda formula \cite{orbitalmagterm1}.
Furthermore, {\axy} is expressed using $\sigma_\mathrm{xy}$ as 
\begin{equation}
  \alpha_{xy}(T,\mu) = -\frac{1}{e}\int d\varepsilon \sigma_{xy}(0,\varepsilon)\frac{\varepsilon - \mu}{T}(-\frac{\partial f}{\partial \varepsilon}).
\end{equation}
Namely, {\axy} emerges as the second derivative of {\Morb} with respect to energy.
\color{black}
This suggests that neither the magnitude nor even the presence of {\color{black}$M_\mathrm{z}^\mathrm{spin}$} is essential for a significant ANE signal.

\color{black}
We have carried out first-principles calculations to confirm that in-plane spin magnetization induces finite out-of-plane orbital magnetization through the spin-orbit coupling, thereby generating finite {\axy}. 
Band structures, momentum distribution of $\Omega_\mathrm{z}$ and {\Morb}, and {\axy} are calculated for two in-plane spin magnetization states $\bm{M}^\mathrm{spin}\parallel$ [11$\bar{\text{2}}$] $(x)$ and $\bm{M}^\mathrm{spin}\parallel$ [$\bar{\text{1}}$10] $(y)$ of the (111) {\SRO} film (see Supplementary Note 8 for details). 
Figures 4\textbf{c} and 4\textbf{d} present the momentum distribution of $\Omega_\mathrm{z}$ at the Fermi surface and $M_\mathrm{z}^\mathrm{orb}$ for the $\bm{M}^\mathrm{spin}\parallel$ [11$\bar{\text{2}}$] state. 
The magnetic point group $2'/m'$ in this state contains inversion symmetry $P$ but no unitary $C_2$ rotations along in-plane directions or vertical mirror operations.
Instead, only antiunitary operations $C_{2y}T$ and $\sigma_y T$ are preserved, which allows uncancelled $\Omega_\mathrm{z}$ distribution at the Fermi surface due to $\sigma_y T\Omega_z(k_x,k_y,k_z) = +\Omega_z(k_x,-k_y,k_z)$ and $C_{2y} T\Omega_z(k_x,k_y,k_z) = +\Omega_z(-k_x,k_y,-k_z)$.
This also gives rise to finite {\axy} as well as $M_\mathrm{z}^\mathrm{orb}$.
The discovered in-plane ANE can thus be understood from the above symmetry arguments governing the emergence of finite {\axy} and out-of-plane orbital magnetization {\Morb} induced by in-plane spin magnetization.

In fact, the calculated transverse thermoelectric conductivity is {$|\alpha_\mathrm{xy}| = \SI{0.0049}{A/Km}$} for $\bm{M}^\mathrm{spin} \parallel$ [11$\bar{\text{2}}$] $(x)$, in contrast to zero for $\bm{M}^\mathrm{spin} \parallel$ [$\bar{\text{1}}$10] $(y)$.
In comparison, the experimentally obtained values are {$\alpha_\mathrm{xy} = \SI{-0.039}{A/Km}$} for $\bm{M}^\mathrm{spin} \parallel$ [11$\bar{\text{2}}$] and zero for $\bm{M}^\mathrm{spin} \parallel$ [$\bar{\text{1}}$10], consistent with the symmetry arguments discussed above.
Also, the relative orbital magnetization {\dMorb} calculated by integrating over the energy range from $-3.5\,\mathrm{eV}$ to $0\,\mathrm{eV}$ is {$\Delta M_\mathrm{z}^\mathrm{orb} (\mu) = \num{-0.013}\, \mu_\mathrm{B} /V_\mathrm{cell}$} for $\bm{M}^\mathrm{spin} \parallel$ [11$\bar{\text{2}}$], in contrast to zero for $\bm{M}^\mathrm{spin} \parallel$ [$\bar{\text{1}}$10].
These calculations support that finite ANE associated with out-of-plane orbital magnetization emerges even when the spin magnetization is lying in in-plane directions of the trigonally distorted {\SRO} films.
\color{black}

While here we demonstrate the observation of in-plane ANE in the symmetry-tailored {\SRO} films, its emergence can be expected in a much broader class of materials. 
In particular, materials that exhibit large intrinsic ANE derived from Berry curvature hot spots or Weyl points \cite{ANEexp4,ANEexp3,ANEexp5} are promising candidates, if the symmetry requirements are satisfied on the temperature gradient-electric voltage plane. 
{\color{black} Furthermore, lifting the orthogonality condition of ANE is expected to enable more flexible thermoelectric device design.
For example, in thermopile structures based on conventional ANE, a temperature gradient is applied perpendicular to the plane, and the electromotive force is generated along the stripe direction, requiring in-plane spin magnetization perpendicular to the stripes as shown in Fig. 5\textbf{a}.
However, this is an unfavorable orientation from the viewpoint of shape magnetic anisotropy, and consequently, device fabrication becomes more challenging as the stripes become narrower. 
In contrast, in-plane ANE enables considerable electromotive force when spin magnetization is aligned along the stripes or even perpendicular to the plane, as shown in Fig. 5\textbf{b}.
Therefore, the stripe width in thermopile structures could be further reduced, thereby leading to smaller devices.
Another prominent feature of in-plane ANE is that its sign can be controlled by engineering crystal symmetry. 
Thermopile structures based on bipolar in-plane ANE are expected to be realized through symmetry engineering such as by domain engineering and heterostructuring \cite{2Dengineering}, beyond the conventional Fermi level tuning \cite{ANEthermopile}.}
Namely, the finding of in-plane ANE serves not only to lift the orthogonality restriction but also to provide a new approach to the polarity control by harnessing the in-plane degrees of freedom.

\section*{Conclusion}
In summary, we have succeeded in observing {\color{black}in-plane ANE in trigonally distorted {\SRO} ultrathin films, where the spin magnetization lies in the plane defined by the temperature gradient and the induced voltage}. 
Azimuthal field-angle sweeps exhibit the three-fold symmetric ANE signal reflecting the crystal symmetry. 
Polar field-angle sweeps also reveal that ANE spontaneously emerges with coupling to the in-plane spin magnetization, reaching a value approximately \SI{40}{\%} of the conventional ANE under the out-of-plane spin magnetization. 
{\color{black} The observed temperature dependence also indicates that it is indeed related to the in-plane ferromagnetic ordering in {\SRO}.}
These magnetothermoelectric measurements performed for spherical rotations of the magnetic field demonstrate that {\color{black}in-plane ANE} can be induced through the off-diagonal coupling between the in-plane spin magnetization and the out-of-plane orbital magnetization. 
{\color{black} The first-principles calculations also confirm that in-plane spin magnetization induces finite out-of-plane orbital magnetization, thereby generating finite thermoelectric conductivity.}
Further in-plane transverse thermoelectric effects, such as in the anomalous Ettingshausen effect and the thermal Hall effect, can be expected to emerge through this off-diagonal coupling. 
Our findings extend the potential of anomalous Nernst effect especially by lifting the orthogonality restriction among spin magnetization, temperature gradient, and electric voltage, opening to more flexible designs of magnetothermoelectric materials and devices.


\begin{methods}
\noindent \textbf{Epitaxial film growth}

{\SRO} thin films with (111) orientation were fabricated on (111) {\STO} substrates using an Eiko EB-9000S oxide molecular beam epitaxy system equipped with a semiconductor-laser heating unit \cite{SRO113MBE,SRO327MBE}. Before the film growth, the substrates were annealed at \SI{870}{\degreeCelsius} to prepare the surface. The {\SRO} films were subsequently grown at \SI{650}{\degreeCelsius} with supplying 4N Sr from a Knudsen cell, 3N5 Ru from an electron beam evaporator, and an O$_3$ (60\%) + O$_2$ (40\%) mixture gas from a Meidensya MPOG-RDE01C ozone generator. The film thickness was designed to be approximately \SI{4}{nm}.

\noindent \textbf{Magnetotransport measurements}

Longitudinal resistivity {\rhoxx} and Hall resistivity {\rhoyx} were measured on a {\SRO} Hall bar device, which was fabricated on the {\STO} substrate with typical dimensions of $3.5\times 1.5\times 0.3$ mm$^3$, as shown in Fig. 2\textbf{a} (see also Supplementary Fig. 1). The magnetotransport measurements were performed using a conventional low-frequency lock-in technique for a standard four-probe method, with an electric current of \SI{5}{\micro A} applied along the [11$\bar{\text{2}}$] crystalline direction. Angle-dependent magnetotransport was measured up to 9 T in a Cryomagnetics cryostat system equipped with a superconducting magnet and a sample rotator probe. The magnetic field direction was varied either within the (111) plane, with the azimuthal angle $\varphi$ measured from [11$\bar{\text{2}}$], or from the out-of-plane [111] direction toward an in-plane [11$\bar{\text{2}}$] one, defined by the polar angle $\theta$ measured from [111].

\noindent \textbf{Magnetothermoelectric measurements}

The Nernst measurements up to 9 T were performed on the same {\SRO} Hall bar device immediately after the magnetotransport measurements, with the magnetic field rotated under the same angular definitions of $\varphi$ and $\theta$. 
To generate a thermal gradient along the [11$\bar{\text{2}}$] crystalline direction, a \SI{1}{k\ohm} chip resistor was placed at one silver-pasted end of the sample as a heater, powered by a Keithley 2450 Source Measure Unit, while the other end was kept in contact with a heat bath. 
The longitudinal thermal gradient {\nablaxT} was determined by monitoring temperature difference {\dxT} between two points approximately 1.7 mm apart using commercial NiCr-CuNi thermocouples. 
The transverse voltage {\Vy}, simultaneously measured using a Keithley 2182A Nanovoltmeter, is confirmed to increase linearly with {\color{black} {\dxT} (see Supplementary Fig. 2).
The corresponding transverse electric field $E_\mathrm{y}$ is defined as {\Vy} divided by the distance between the voltage terminals, and then the Nernst signal is obtained as $S_\mathrm{yx} = E_\mathrm{y}/ (-\nabla _\mathrm{x}T)$} after subtracting the longitudinal component symmetric with respect to magnetic field. 
For the in-plane field sweep and rotation measurements, the one-fold symmetric component with respect to the in-plane field rotation is further subtracted (see Supplementary Note {\color{black}6} for details). 
For the actual measurements of {\Syx} under the field, {\Vy} was measured continuously for {\color{black} fixed temperature differences of {\dxT $= \num{0}, \num{6},$ and $\SI{12}{K}$}} during the field sweep or rotation. 
In the case of {\SyxzeroT}, on the other hand, the relation between {\Vy} and {\dxT} was measured for each poling field direction.

\noindent \textbf{First-principles calculation}

\color{black}{
First-principles calculations for trigonally distorted {\SRO} were carried out within the DFT+U framework using the generalized gradient approximation in the Perdew-Burke-Ernzerhof (PBE) form \cite{cal1, cal2}. 
The calculations employed the projector-augmented-wave (PAW) method \cite{cal3} as implemented in the Vienna ab initio simulation package (VASP) \cite{cal4, cal5}. 
A plane-wave cutoff energy of \SI{550}{eV} and a $\Gamma$-centered $13\times 13\times 13$ k-point mesh were used for the Brillouin-zone sampling. 
Electron correlation effects on the Ru $d$ orbitals were treated using the rotationally invariant DFT+U scheme \cite{cal6} with $U = \SI{2.5}{eV}$ and $J = \SI{0.5}{eV}$. 
We adopted the experimental structural parameters of trigonally distorted bulk {\SRO}, with lattice constants $a = b = c = \num{3.91988}\,\text{\AA}$ and angles $\alpha = \beta = \gamma = 89.5658^\circ$. 
The band structures were generated with VASPKIT \cite{cal7}, while crystal structures were visualized using VESTA \cite{cal8}. 
Based on the resulting DFT+U band structures, a Wannier tight-binding model was constructed from 40 orbitals, including Ru $p$ and $d$ orbitals as well as O $s$ and $p$ orbitals \cite{cal9}. 
The intrinsic anomalous Hall conductivity and the corresponding thermoelectric conductivity were then calculated from the Berry-curvature contribution through Brillouin-zone integration using Wannier interpolation \cite{cal9, cal10}. 
To ensure numerical convergence, a dense $500 \times 500 \times 500$ k-point mesh was used in the integration. 
In addition, the Wannier Hamiltonian was symmetrized with Wannsymm \cite{cal11} according to the relevant magnetic point group, enabling accurate calculations of the k-resolved Berry curvature and orbital magnetization, which were analyzed using WannierTools \cite{cal12}. 
Note that only the itinerant part of the orbital magnetization is calculated through chemical potential integration of the calculated anomalous Hall conductivity.
}

\end{methods}

\noindent \textbf{Data availability}
\noindent The data supporting the plots within the paper and its Supplementary information file are available from the corresponding author upon reasonable request.

\newpage
\renewcommand{\baselinestretch}{1}
\renewcommand{\baselinestretch}{1.5}\normalsize

\begin{addendum}
 \item This work was supported by JSPS KAKENHI Grant Numbers JP23K13666, JP23K03275, JP24H01614, JP24H01654, JP25H00841, {\color{black}JP25H01252, JP26H00594, and JP26H00646} from MEXT, Japan, by JST FOREST Program Grant Number JPMJFR202N and PRESTO Program Grant Number JPMJPR2452, by Toyota Riken Rising Fellow Program funded by Toyota Physical and Chemical Research Institute, Japan, and by STAR Award funded by the Tokyo Tech Fund, Japan.
 \item[Author contribution] M.U. conceived the project and designed the experiments. T.Y. performed transport measurements with S.N. and M.K. H.K. grew films with {\color{black}N.T. and} Y.M. {\color{black}M.-C.J. performed first-principles calculations with R.A.} T.Y. and S.N. analyzed the data and T.Y. and M.U. wrote the manuscript with input from all authors. H.I. jointly discussed the results. All authors have approved the final version of the manuscript.
 \item[Competing Interests] The authors declare no competing interests.
 \item[Additional information] Supplementary information is available in the online version. Reprints and permissions information is available online at www.nature.com/reprints. Correspondence and requests for materials should be addressed to M.U. (email: m.uchida@phys.sci.isct.ac.jp).
\end{addendum}

\newpage
\renewcommand{\baselinestretch}{1.3}\normalsize
\begin{figure}
\begin{center}
\includegraphics[width=10cm]{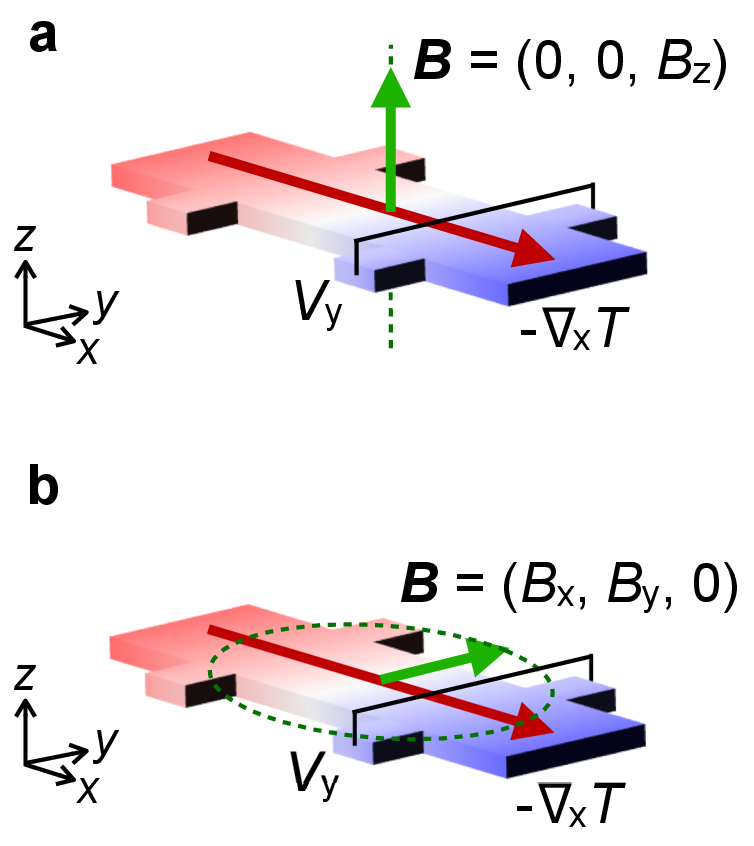}
\caption{
\textbf{Out-of-plane and in-plane anomalous Nernst effects.}
\textbf{a}, Schematic illustration of the configuration for the conventional anomalous Nernst effect under the magnetic field, $\bm{B}$. Anomalous Nernst voltage {\Vy} is generated perpendicular to both the temperature gradient {$-$\nablaxT} and the out-of-plane field component $B_\mathrm{z}$. \textbf{b}, Schematic of the in-plane anomalous Nernst effect. {\Vy} transverse to {$-$\nablaxT} can be induced by the in-plane field components $B_\mathrm{x}$ and $B_\mathrm{y}$.
}
\end{center}
\end{figure}
\clearpage
\newpage

\begin{figure}
\begin{center}
\includegraphics[width=\linewidth]{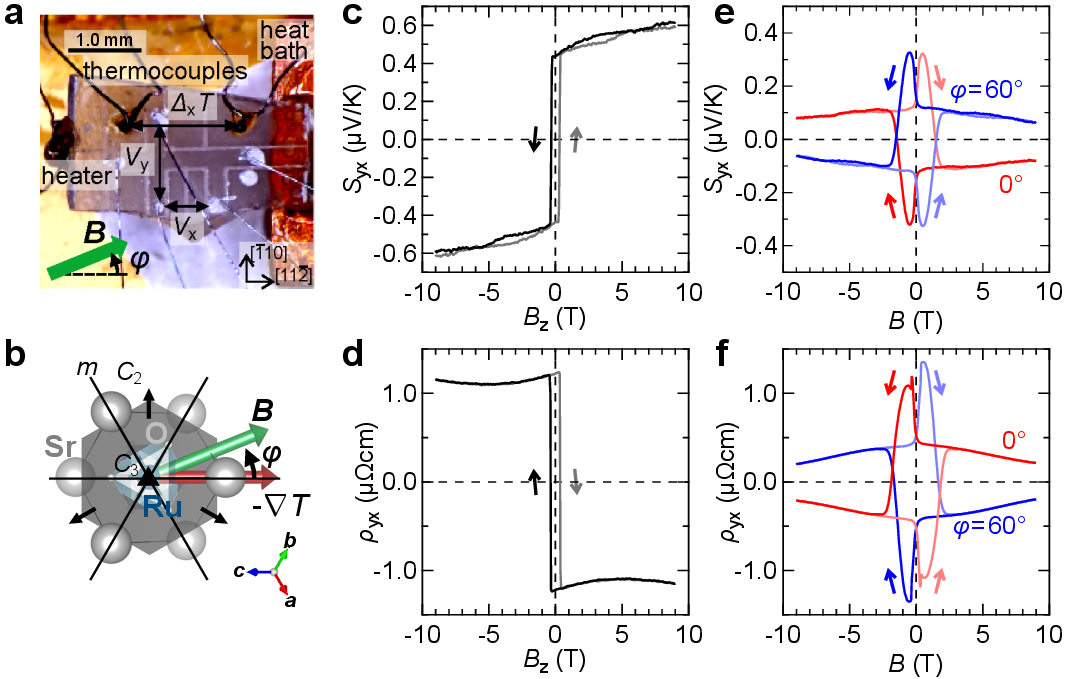}
\caption{
\textbf{In-plane anomalous Nernst effect in a {\SRO} film.}
\textbf{a}, Experimental setup for measuring anomalous Nernst effect of a (111)-oriented {\SRO} film with in-plane magnetic field sweep and rotation. \textbf{b}, Crystal structure of trigonally distorted {\SRO}, shown with fundamental symmetry elements ($C_2$ and $C_3$ rotation axes and mirror planes). \textbf{c}, Nernst signal {\Syx} and \textbf{d}, Hall resistivity {\rhoyx} of the (111) {\SRO} film, taken with conventionally sweeping the magnetic field along the out-of-plane [111] direction at nominal sample temperature of \SI{63}{K}. \textbf{e}, {\Syx} obtained by sweeping the in-plane magnetic field at azimuthal angles $\varphi = 0^\circ$ and $60^\circ$, where $\varphi$ is measured from the [11$\bar{\text{2}}$] direction as shown in \textbf{b}. \textbf{f}, {\rhoyx} taken for the in-plane field sweep at $\varphi = 0^\circ$ and $60^\circ$.
}
\end{center}
\end{figure}
\clearpage
\newpage

\begin{figure}
\begin{center}
\includegraphics[width=0.99\linewidth]{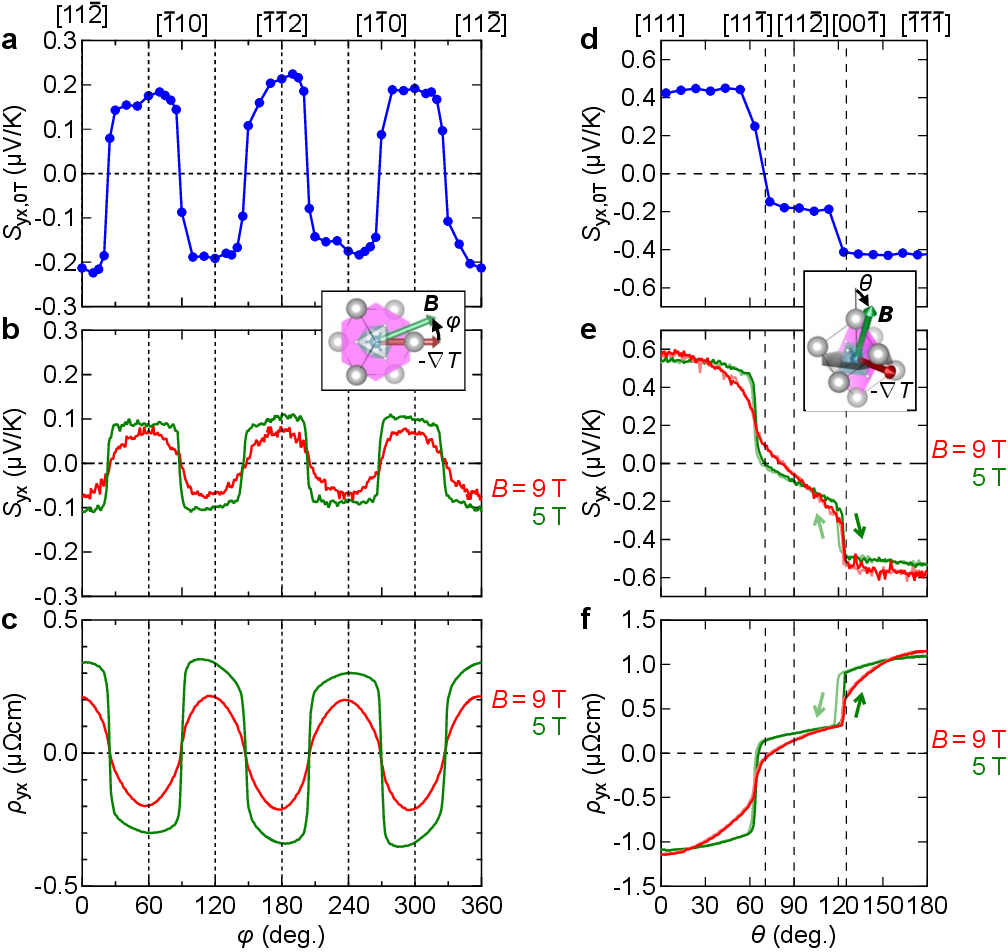}
\caption{
\textbf{\color{black}Azimuthal and polar angle dependence.}
{\color{black}\textbf{a}, Nernst signal at zero magnetic field {\SyxzeroT}, measured after increasing the in-plane magnetic field up to \SI{9}{T}, returning it to \SI{0}{T}, and then setting the heater power to vary {\dxT} at each azimuthal angle $\varphi$. 
\textbf{b}, {\Syx} taken upon continuously rotating the magnetic field within the (111) plane at \SI{5}{T} and \SI{9}{T}. 
\textbf{c}, {\rhoyx} measured with similarly rotating the field within the (111) plane. 
\textbf{b}-\textbf{f}, Corresponding data taken at each polar angle $\theta$ on the $\varphi = 0^\circ$ plane.
All the measurements were performed at nominal sample temperature of \SI{63}{K}.
}}
\end{center}
\end{figure}
\clearpage
\newpage
\begin{figure}
\begin{center}
\includegraphics{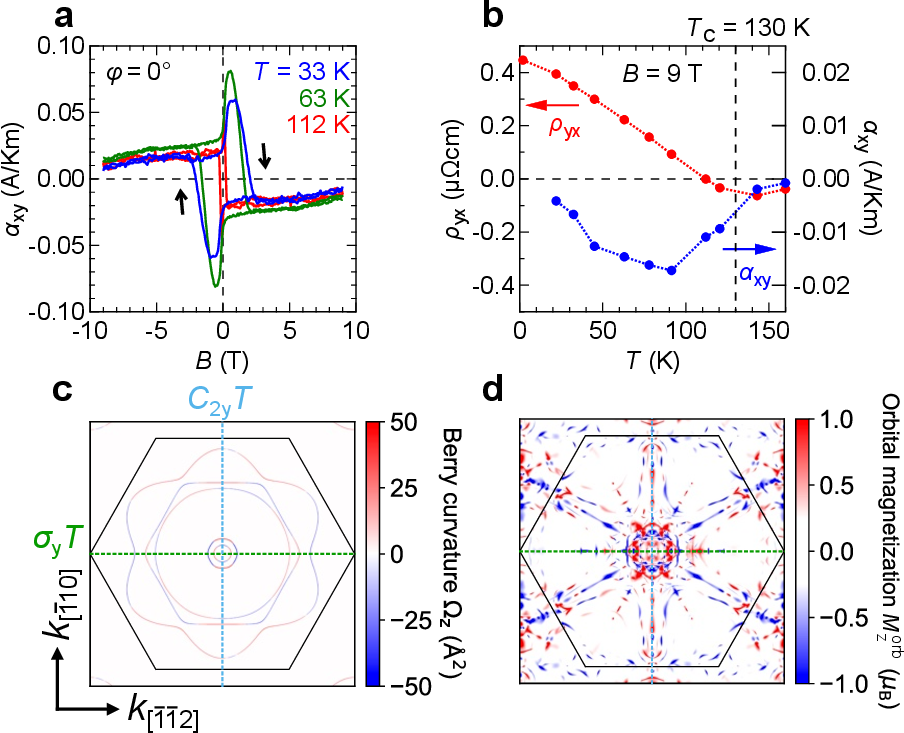}
\caption{
\textbf{\color{black}Temperature dependence of in-plane ANE.}
{\color{black}\textbf{a}, In-plane field dependence of transverse thermoelectric conductivity {\axy} at temperatures of \SI{33}{K}, \SI{63}{K}, and \SI{112}{K} for $\varphi = 0^\circ$, obtained by the equation {$\alpha_{\mathrm{xy}}= (S_{\mathrm{yx}} + \theta_{\mathrm{AHE}}\cdot S_\mathrm{xx})/\rho_{\mathrm{xx}}$}.
\textbf{b}, Temperature dependence of {\rhoyx} and {\axy} under in-plane field of \SI{9}{T} at $\varphi = 0^\circ$.
Momentum distribution of \textbf{c}, out-of-plane Berry curvature $\Omega_\mathrm{z}$ at the Fermi surface 
and \textbf{d}, out-of-plane orbital magnetization $M_\mathrm{z}^\mathrm{orb}$, calculated with the spin magnetization set to the in-plane [11$\bar{\text{2}}$] direction ($\varphi = 0^\circ$ and $\theta = 90^\circ$) and plotted on the (111) plane.
}}
\end{center}
\end{figure}

\clearpage
\newpage
\begin{figure}
\begin{center}
\includegraphics{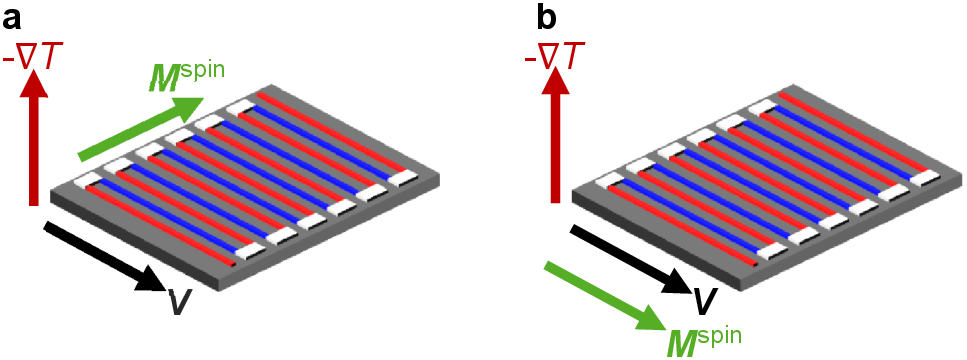}
\caption{
\textbf{\color{black}Thermopile based on the in-plane ANE.}
{\color{black} Schematic illustration of thermopile structures utilizing \textbf{a}, conventional ANE with orthogonality restriction, 
and \textbf{b}, in-plane ANE lifting orthogonality restriction among the temperature gradient, electromotive force, and spin magnetization.
}}
\end{center}
\end{figure}

\clearpage
\end{document}